\RequirePackage{fix-cm}
\documentclass[smallextended]{svjour3}       
\smartqed  
\usepackage{graphicx}
\usepackage{amsmath}
\usepackage{amssymb}
\usepackage{graphicx}
\usepackage{color}

\journalname{Few-Body Systems}

\begin{document}

\title{Valence-quark structure N* resonances from DSEs
%
}


\author{Jorge Segovia}


\institute{J. Segovia (speaker) \at
           Institut de F\'isica d'Altes Energies and Barcelona Institute of Science and Technology, \\ 
           Universitat Aut\`onoma de Barcelona, E-08193 Bellaterra (Barcelona), Spain \\           
           \email{jsegovia@ifae.es}
}

\date{Received: date / Accepted: date}

\maketitle

\begin{abstract}
We present a unified Quantum Chromodynamics (QCD)-based description of elastic and transition electromagnetic form factors involving the nucleon and its resonances. We compare predictions made using a framework built upon a Faddeev equation kernel and interaction vertices that possess QCD-like momentum dependence with results obtained using a confining, symmetry-preserving treatment of a vector$\,\otimes\,$vector contact-interaction in a widely-used leading-order (rainbow-ladder) truncation of QCD's Dyson-Schwinger equations. This comparison explains that the contact-interaction framework produces hard form factors, curtails some quark orbital angular momentum correlations within a baryon, and suppresses two-loop diagrams in the elastic and transition electromagnetic currents. Such defects are rectified in our QCD-based approach and, by contrasting the results obtained for the same observables in both theoretical schemes,
shows those objects which are most sensitive to the momentum dependence of elementary quantities in QCD.
%
\keywords{
Dyson-Schwinger equations \and
elastic and transition electromagnetic form factors \and
nucleon resonances
}
\PACS{
11.80.Jy \and 
13.40Gp  \and 
14.20.Dh      
}
\end{abstract}


\newpage

\section{Introduction}
\label{sec:intro}

A unified description of electromagnetic elastic and transition form factors involving the nucleon and its resonances has acquired very much interest. On the theoretical side, it is via the $Q^2$-evolution of form factors that one gains access to the running of QCD's coupling and masses from the infrared into the ultraviolet~\cite{Cloet:2013gva,Chang:2013nia}. Moreover, QCD-based approaches, able to compute form factors at large photon virtualities, are needed in order to pierce the meson-cloud that, often to a significant extent, screens the dressed-quark core of all baryons~\cite{Tiator:2003uu,Kamano:2013iva}.

On the experimental side, we have witnessed a substantial progress in the extraction of transition electrocouplings, $g_{{\rm v}NN^\ast}$, from meson electroproduction data, obtained primarily with the CLAS detector at the Jefferson Laboratory (JLab). The electrocouplings of all low-lying $N^\ast$ states with mass less-than $1.6\,{\rm GeV}$ have been determined via independent analyses of $\pi^+ n$, $\pi^0p$ and $\pi^+ \pi^- p$ exclusive channels~\cite{Agashe:2014kda,Mokeev:2012vsa}; and preliminary results for the $g_{{\rm v}NN^\ast}$ electrocouplings of most high-lying $N^\ast$ states with masses below $1.8\,{\rm GeV}$ have also been obtained from CLAS meson electroproduction data~\cite{Aznauryan:2012ba,Mokeev:2013kka}.

It is within the context just described that we have performed a simultaneous treatment of electromagnetic elastic and transition form factors involving the Nucleon, Delta and Roper baryons in Refs.~\cite{Segovia:2013rca,Segovia:2013uga,Segovia:2014aza,Segovia:2015ufa,Segovia:2015hra,Xu:2015kta,Segovia:2016zyc}. In order to address the issue of charting the behaviour of the running coupling and masses in the strong interaction sector of the Standard Model, we use a Dyson-Schwinger equations (DSEs) based approach~\cite{Chang:2011vu,Bashir:2012fs,Cloet:2013jya} and compare results between a QCD-based framework and a confining, symmetry-preserving treatment of a vector$\,\otimes\,$vector contact interaction.


\vspace*{-0.50cm}
\section{Baryon structure}
\label{sec:Baryons}

Dynamical chiral symmetry breaking (DCSB) is a theoretically-established feature of QCD and the most important mass generating mechanism for visible matter in the Universe, being responsible for approximately $98\%$ of the proton's mass. A fundamental expression of DCSB is the behaviour of the quark mass-function, $M(p)$. This appears in the dressed-quark propagator which may be obtained as a solution to the most famous and simple QCD's Dyson-Schwinger equation: the gap equation~\cite{Cloet:2013jya}. The nontrivial character of the mass function arises primarily because a dense cloud of gluons comes to clothe a low-momentum quark. It explains how an almost-massless parton-like quark at high energies transforms, at low energies, into a constituent-like quark with an effective mass of around $350\,{\rm MeV}$.

DCSB ensures the existence of nearly-massless pseudo-Goldstone modes (pions). Another equally important consequence of DCSB is less well known. Namely, any interaction capable of creating pseudo-Goldstone modes as bound-states of a light dressed-quark and -antiquark, and reproducing the measured value of their leptonic decay constants, will necessarily also generate strong colour-antitriplet correlations between any two dressed quarks contained within a baryon. Although a rigorous proof within QCD cannot be claimed, this assertion is based upon an accumulated body of evidence, gathered in two decades of studying two- and three-body bound-state problems in hadron physics (the interested reader is referred to the discussion in Ref.~\cite{Segovia:2015ufa} and to Refs.~[21-35] cited therein). No realistic counter examples are known; and the existence of such diquark correlations is also supported by simulations of lattice QCD~\cite{Alexandrou:2006cq,Babich:2007ah}.

\begin{figure}[!t]
\begin{center}
\hspace*{0.50cm}
\includegraphics[clip,width=0.40\textwidth,height=0.18\textheight]
{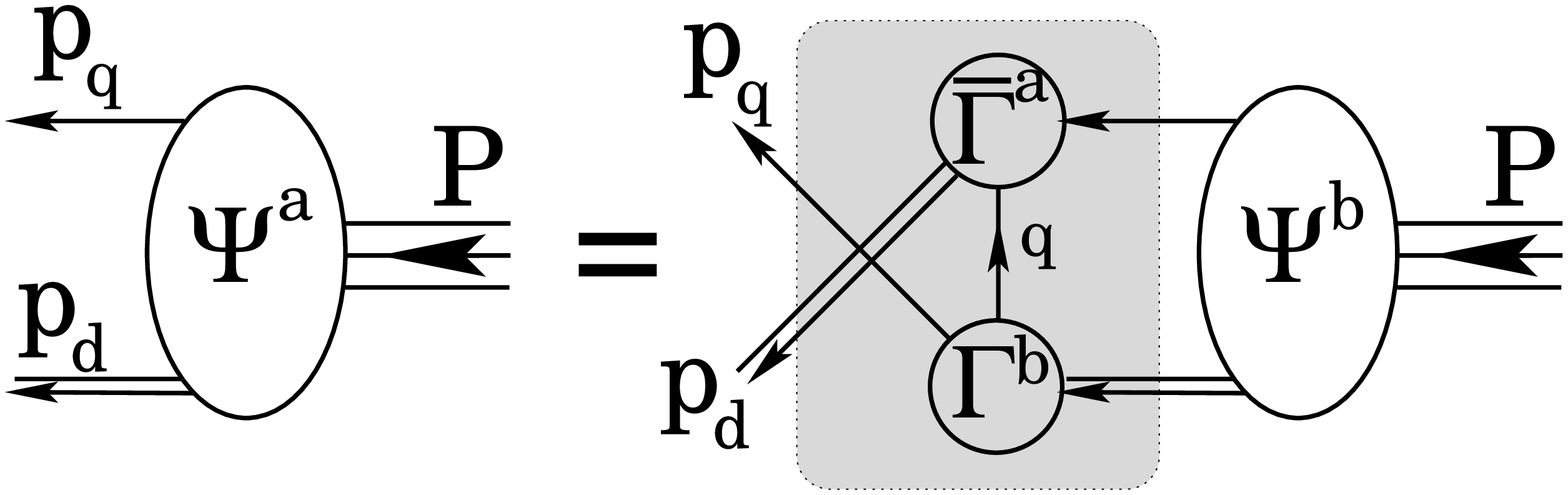} 
\hspace*{1.00cm}
\includegraphics[clip,width=0.40\textwidth,height=0.20\textheight]
{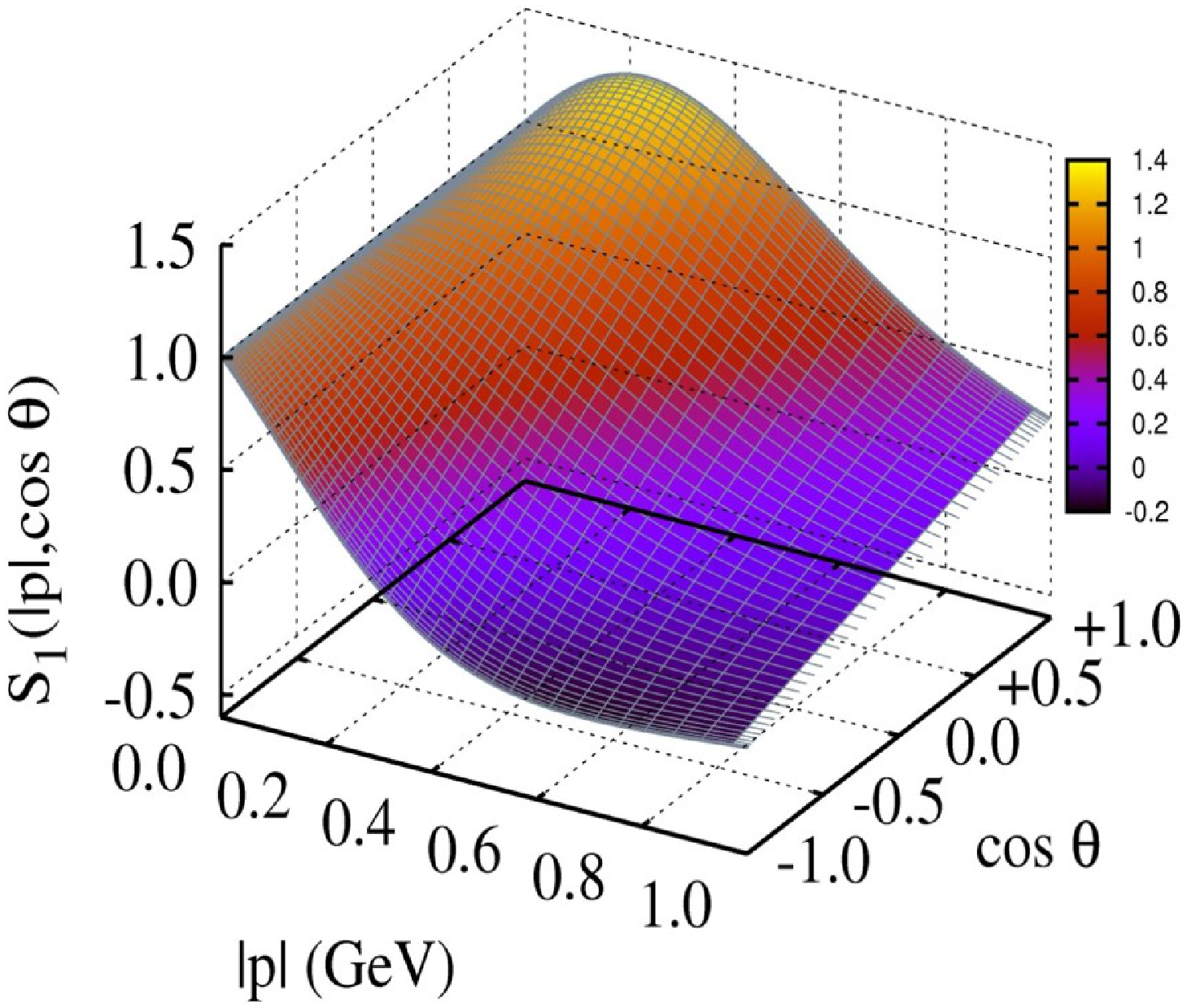}
\caption{\label{fig:Faddeev} {\it Left panel:} Poincar\'e covariant Faddeev equation. $\Psi$ is the Faddeev amplitude for a baryon of total momentum $P= p_q + p_d$, where $p_{q,d}$ are, respectively, the momenta of the quark and diquark within the bound-state. The shaded area demarcates the Faddeev equation kernel: {\it single line}, dressed-quark propagator; $\Gamma$, diquark correlation amplitude; and {\it double line}, diquark propagator.
{\it Right panel:} Dominant piece in the nucleon's eight-component Poincar\'e-covariant Faddeev amplitude: $S_1(|p|,\cos\theta)$. In the nucleon rest frame, this term describes that piece of the quark--scalar-diquark relative momentum correlation which possesses zero intrinsic quark-diquark orbital angular momentum, i.e. $L=0$, before the propagator lines are reattached to form the Faddeev wave function. Referring to Fig.~\ref{fig:Faddeev}, $p= P/3-p_q$ and $\cos\theta = p\cdot P/\sqrt{p^2 P^2}$. The amplitude is normalized such that its $U_0$ Chebyshev moment is unity at $|p|=0$.
}
\vspace*{-0.50cm}
\end{center}
\end{figure}

The existence of diquark correlations considerably simplifies analyses of the three valence-quark scattering problem because it reduces that task to solving a Poincar\'e covariant Faddeev equation depicted in the left panel of Fig.~\ref{fig:Faddeev}. Two main contributions appear in the baryon's binding energy. One part is expressed in the formation of tight diquark correlations. That is augmented, however, by attraction generated by the quark exchange depicted in the shaded area of the left panel of Fig.~\ref{fig:Faddeev}. This  exchange ensures that diquark correlations within the baryon are fully dynamical: no quark holds a special place because each one participates in all diquarks to the fullest extent allowed by its quantum numbers. The continual rearrangement of the quarks guarantees, \emph{inter} \emph{alia}, that the baryon's dressed-quark wave function complies with Pauli statistics.

The quark$+$diquark structure of the nucleon is elucidated in the right panel of Fig.~\ref{fig:Faddeev}, which depicts the leading component of its Faddeev amplitude: with the notation of Ref.~\cite{Segovia:2014aza}, $S_1(|p|,\cos\theta)$, computed using the Faddeev kernel described therein. This function describes a piece of the quark$+$scalar-diquark relative momentum correlation. Notably, in this solution of a realistic Faddeev equation there is strong variation with respect to both arguments. Support is concentrated in the forward direction, $\cos\theta >0$, so that alignment of $p$ and $P$ is favored; and the amplitude peaks at $(|p|\simeq M_N/6,\cos\theta=1)$, whereat $p_q \approx P/2 \approx p_d$ and hence the \emph{natural} relative momentum is zero. In the anti-parallel direction, $\cos\theta<0$, support is concentrated at $|p|=0$.


\vspace*{-0.50cm}
\section{The \mbox{\boldmath $\gamma^\ast N(940) \to \Delta(1232)$} Transition}
\label{sec:FFnucdel}

The electromagnetic $\gamma^{\ast}N\to \Delta$ transition can be described in function of three Poincar\'e-invariant form factors~\cite{Jones:1972ky}: magnetic-dipole, $G_{M}^{\ast}$, electric quadrupole, $G_{E}^{\ast}$, and Coulomb (longitudinal) quadrupole, $G_{C}^{\ast}$; that can be extracted in the Dyson-Schwinger approach by a sensible set of projection operators~\cite{Eichmann:2011aa}. The following ratios: $R_{\rm EM} = -G_E^{\ast}/G_M^{\ast}$ and $R_{\rm SM} = - (|\vec{Q}|/2 m_\Delta) G_C^{\ast}/G_M^{\ast}$, are often considered because they can be read as measures of the deformation of the hadrons involved in the reaction and how such deformation influences the structure of the transition current.

\begin{figure}[!t]
\begin{center}
\begin{tabular}{ll}
\includegraphics[clip,height=0.20\textheight,width=0.43\textwidth]{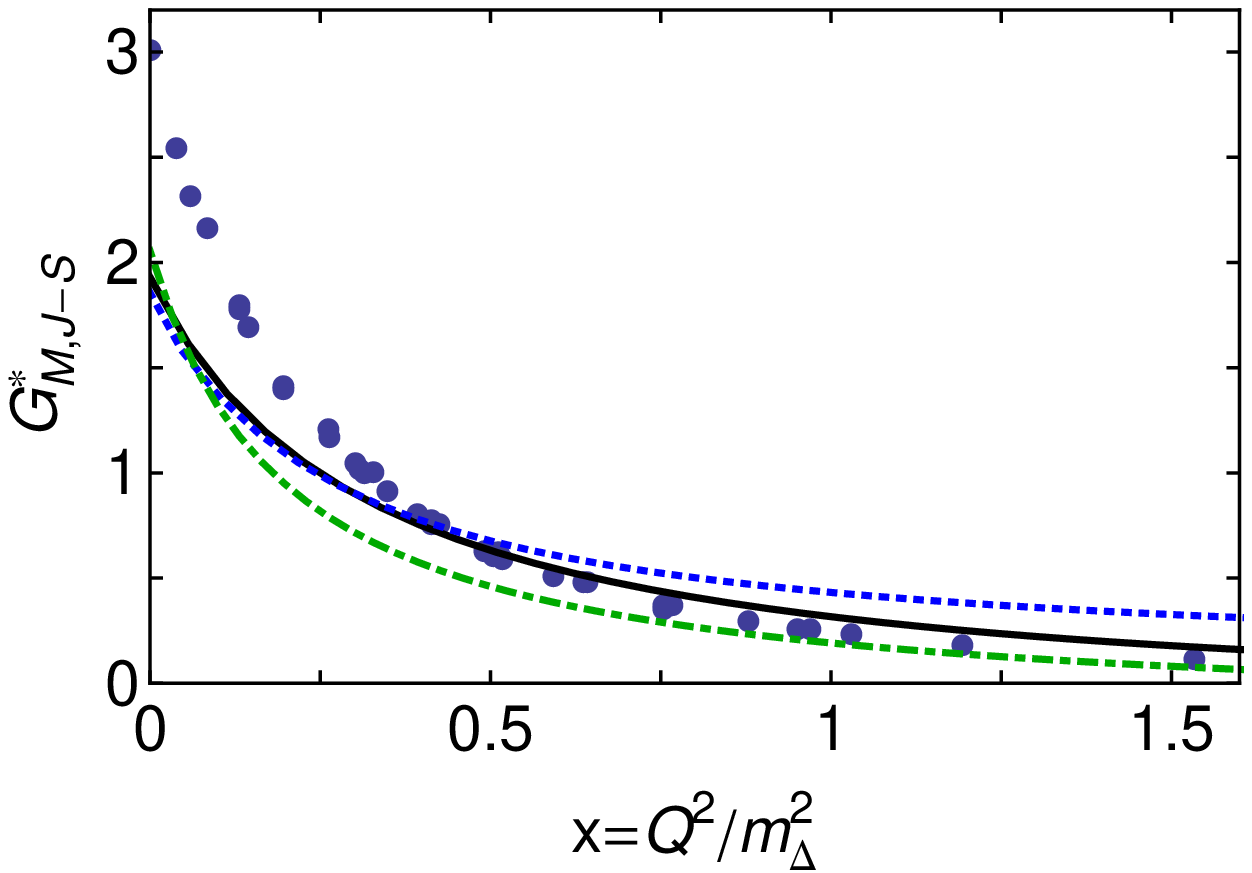}
& 
\includegraphics[clip,height=0.2025\textheight,width=0.45\textwidth]{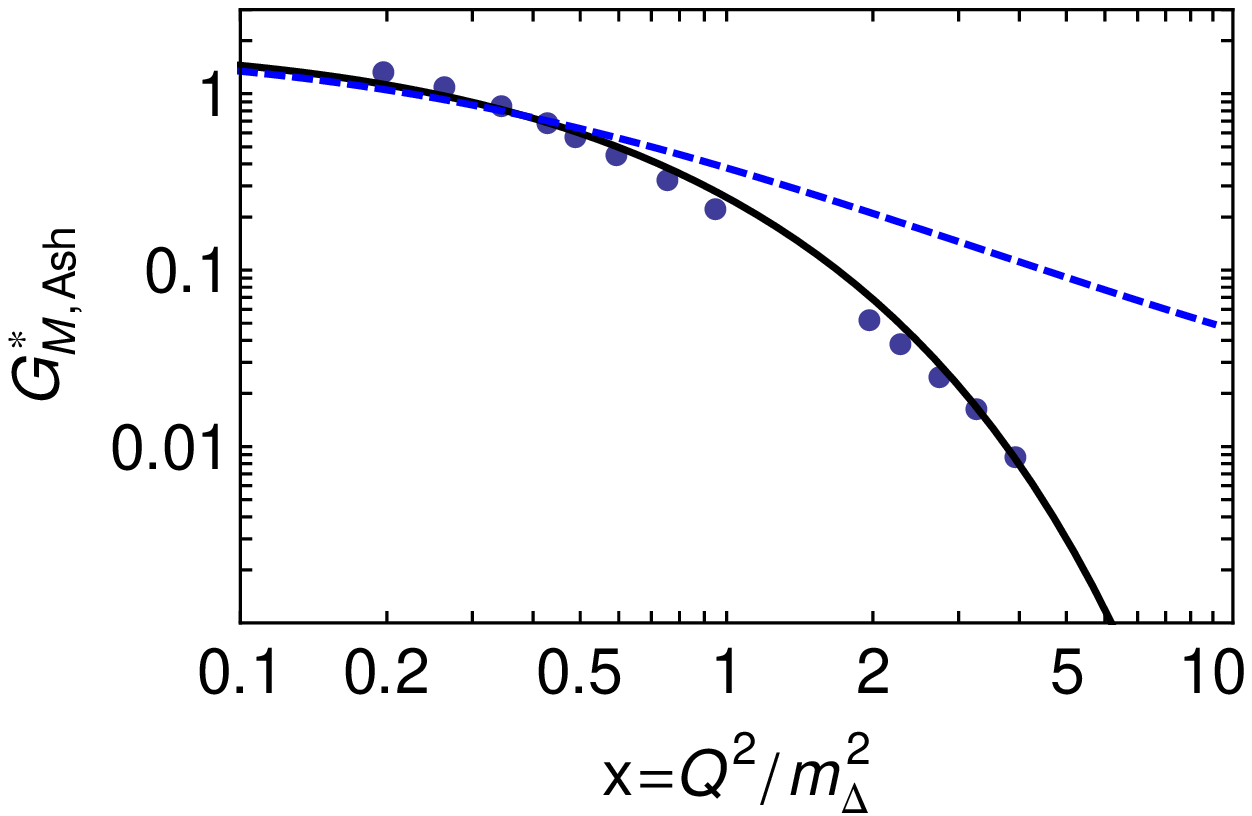} \\
\hspace*{-0.50cm}
\includegraphics[clip,height=0.20\textheight,width=0.455\textwidth]{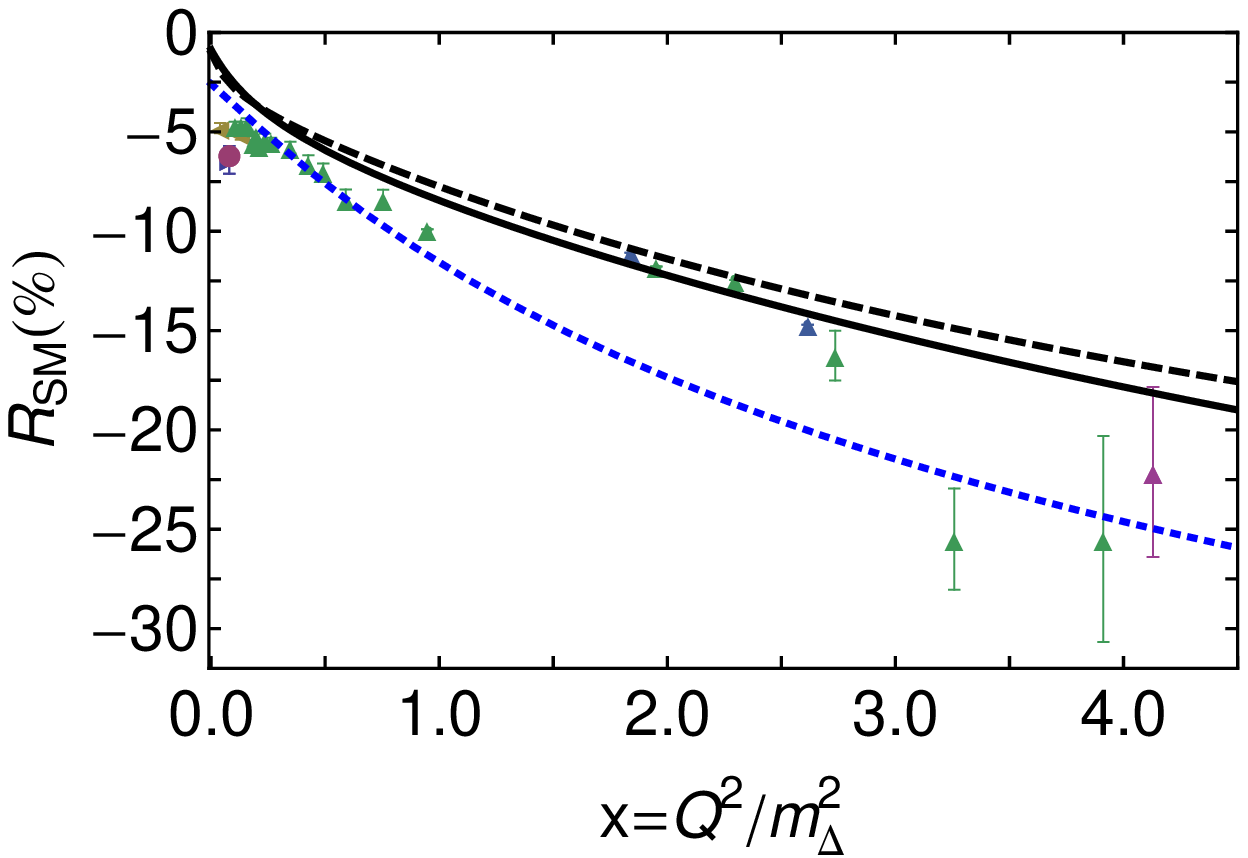}
& 
\includegraphics[clip,height=0.20\textheight,width=0.45\textwidth]{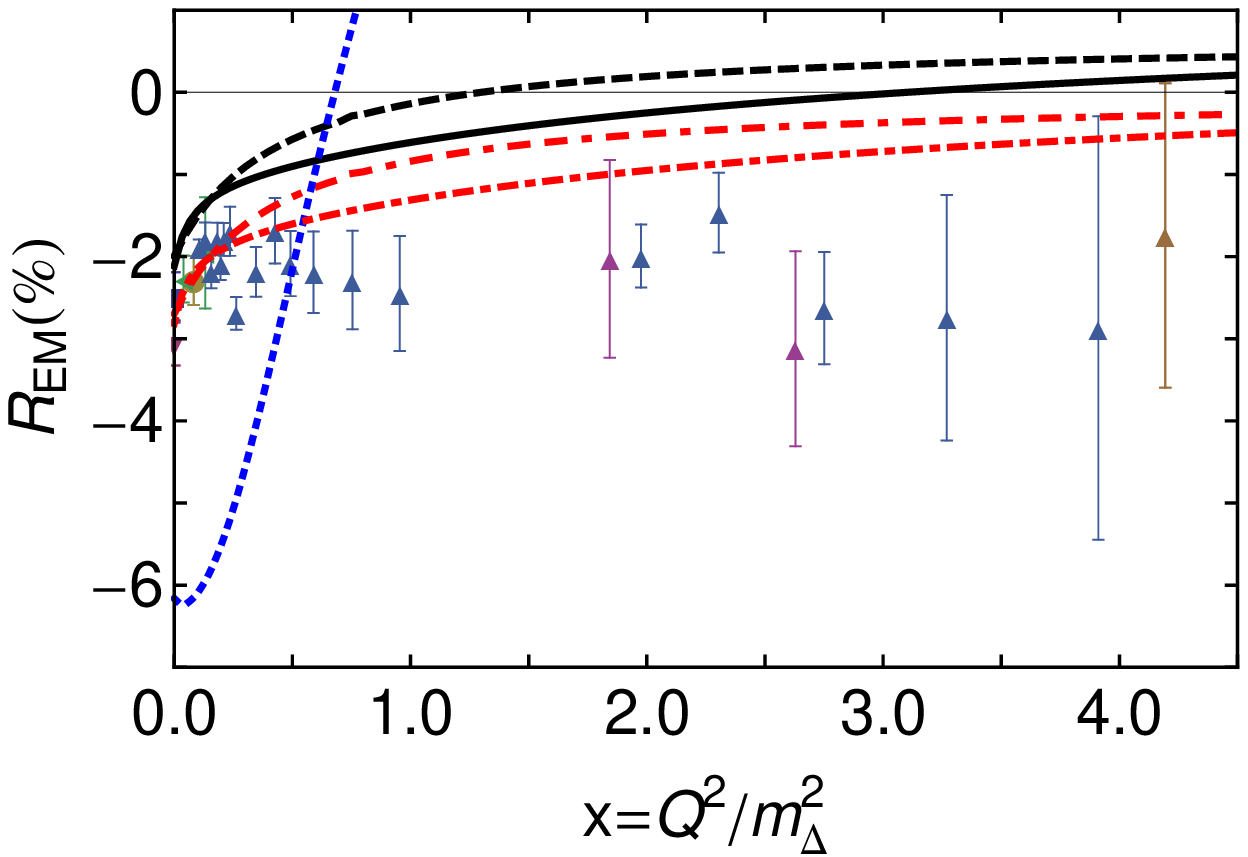}
\end{tabular}
\caption{\label{fig:NucDel} 
\emph{Upper-left panel} -- $G_{M,J-S}^{\ast}$ result obtained with QCD-based interaction (solid, black) and with contact-interaction (CI) (dotted, blue); The green dot-dashed curve is the dressed-quark core contribution inferred using SL-model~\protect\cite{JuliaDiaz:2006xt}.
\emph{Upper-right panel} -- $G_{M,Ash}^{\ast}$ result obtained with QCD-based interaction (solid, black) and with CI (dotted, blue).
\emph{Lower-left panel} -- $R_{SM}$ prediction of QCD-based kernel including dressed-quark anomalous magnetic moment (DqAMM) (black, solid), non-including DqAMM (black, dashed), and CI result (dotted, blue).
\emph{Lower-right panel} -- $R_{EM}$ prediction obtained with QCD-kindred framework (solid, black); same input but without DqAMM (dashed, black); these results renormalized (by a factor of $1.34$) to agree with experiment at $x=0$ (dot-dashed, red - zero at $x\approx 14$; and dot-dash-dashed, red, zero at $x\approx 6$); and CI result (dotted, blue).
The data in the panels are from references that can be found in~\protect\cite{Segovia:2014aza}.
}
\vspace*{-0.5cm}
\end{center}
\end{figure}

The upper-left panel of Fig.~\ref{fig:NucDel} displays the magnetic transition form factor in the Jones-Scadron convention. Our prediction obtained with a QCD-based kernel agrees with the data on $x\gtrsim 0.4$, and a similar conclusion can be inferred from the contact interaction result. On the other hand, both curves disagree markedly with the data at infrared momenta. This is explained by the similarity between these predictions and the bare result determined using the Sato-Lee (SL) dynamical meson-exchange model~\cite{JuliaDiaz:2006xt}. The SL result supports a view that the discrepancy owes to omission of meson-cloud effects in the DSEs' computations.

Presentations of the experimental data associated with the magnetic transition form factor typically use the Ash convention. This comparison is depicted in the upper-right panel of Fig.~\ref{fig:NucDel}. One can see that the difference between form factors obtained with the QCD-kindred and CI frameworks increases with the transfer momentum. Moreover, the normalized QCD-kindred curve is in fair agreement with the data, indicating that the Ash form factor falls unexpectedly rapidly mainly for two reasons. First: meson-cloud effects provide up-to $35\%$ of the form factor for $x \lesssim 2$; these contributions are very soft and hence disappear quickly. Second: the additional kinematic factor $\sim 1/\sqrt{Q^2}$ that appears between Ash and Jones-Scadron conventions and provides material damping for $x\gtrsim 2$ (see Ref.~\cite{Segovia:2014aza} for details on this aspect).

Our predictions for the electromagnetic ratios are depicted in the lower panels of Fig.~\ref{fig:NucDel}. The lower-left panel displays the Coulomb quadrupole ratio. Both the prediction obtained with QCD-like propagators and vertices and the contact-interaction result are broadly consistent with available data. This shows that even a contact-interaction can produce correlations between dressed-quarks within Faddeev wave-functions and related features in the current that are comparable in size with those observed empirically. Moreover, suppressing the dressed-quark anomalous magnetic moment (DqAMM) in the transition current has little impact. These remarks highlight that $R_{SM}$ is not particularly sensitive to details of the Faddeev kernel and transition current.

The differences between the curves displayed in the lower-right panel in Fig.~\ref{fig:NucDel} show that $R_{\rm EM}$ is a particularly sensitive measure of diquark and orbital angular momentum correlations. The contact-interaction result is inconsistent with data, possessing a zero that appears at a rather small value of $x$. On the other hand, predictions obtained with QCD-like propagators and vertices can be viable. We have presented four variants, which differ primarily in the location of the zero that is a feature of this ratio in all cases we have considered. The inclusion of a DqAMM shifts the zero to a larger value of $x$. Given the uniformly small value of this ratio and its sensitivity to the DqAMM, we judge that meson-cloud affects must play a large role on the entire domain that is currently accessible to experiment.


\vspace*{-0.50cm}
\section{The \mbox{\boldmath $\gamma^\ast N(940) \to N(1440)$} Transition}
\label{sec:Roper}

The Roper resonance is at heart the nucleon's first radial excitation, consisting of a well-defined dressed-quark core augmented by a meson cloud that reduces its (Breit-Wigner) mass by approximately 20\%~\cite{Suzuki:2009nj,Segovia:2015hra,Roberts:2016dnb,Aznauryan:2016wwm,Burkert:2017djo}. As part of this explanation, a meson-cloud obscures the dressed-quark core from long-wavelength probes, but that core is revealed to probes with $x_N = Q^2/m_N^2 \gtrsim 3$. Here we summarize the results presented in Ref.~\cite{Segovia:2015hra} about the nucleon-Roper transition form factors. We also show their flavour-separated versions~\cite{Segovia:2016zyc}; since experiments have already yielded precise information on proton-Roper transition form factors~\cite{Dugger:2009pn,Aznauryan:2009mx,Aznauryan:2011qj,Mokeev:2012vsa,Mokeev:2015lda,Burkert:2016dxc}, these predictions could be validated following electroproduction experiments on (bound-)\,neutron targets.

\begin{figure}[!t]
\centerline{%
\includegraphics[clip,width=0.475\textwidth]{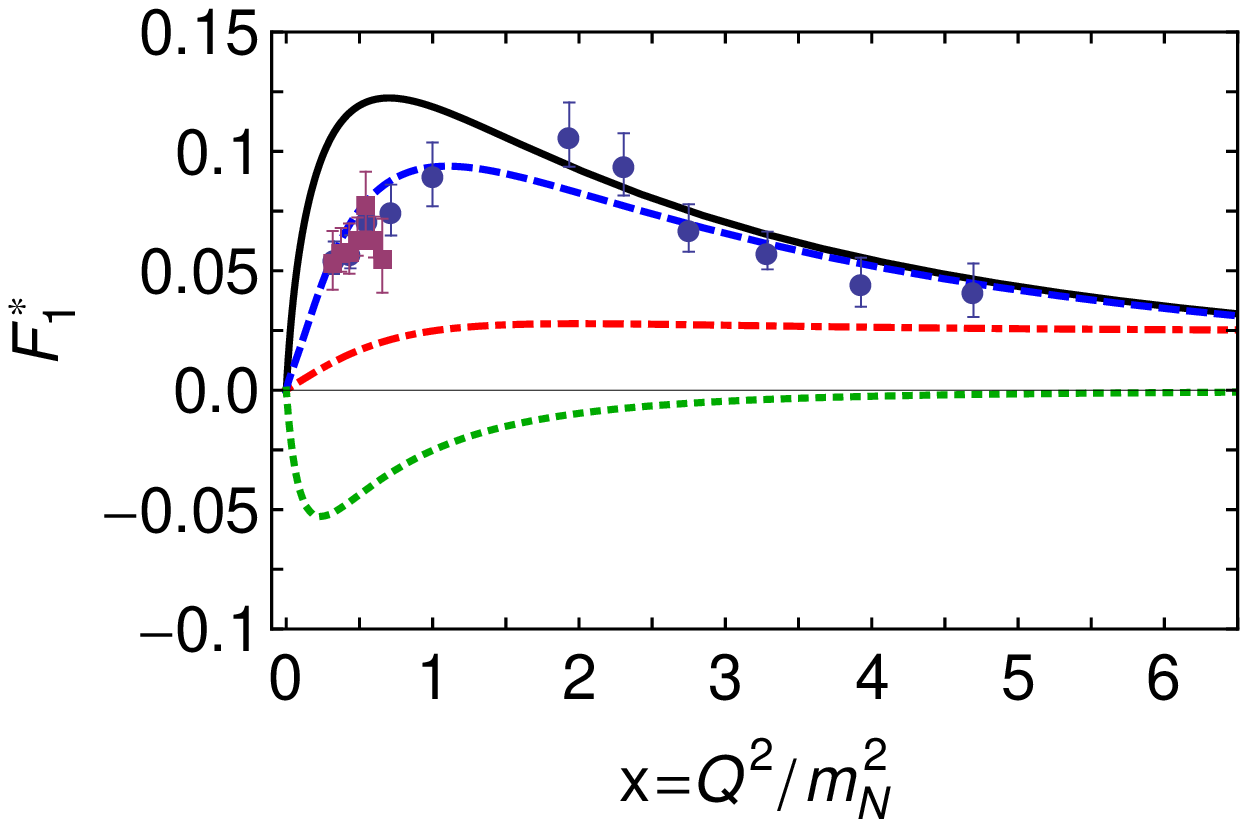}
\hspace*{0.25cm}
\includegraphics[clip,width=0.475\textwidth]{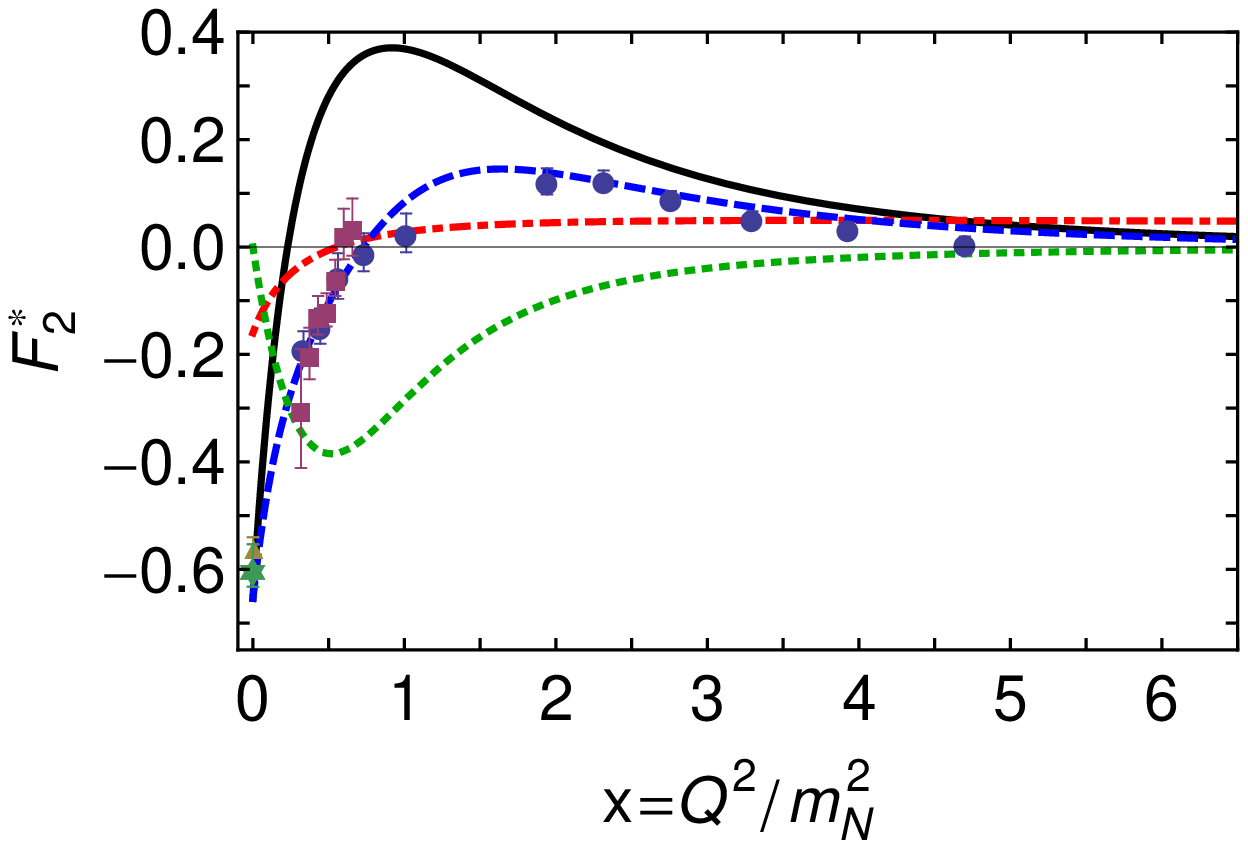}
}
\caption{\label{fig:NucRop_v2} \emph{Left} -- Dirac transition form factor, $F_{1}^{\ast}(x)$, $x=Q^2/m_N^2$. Solid (black) curve, QCD-kindred prediction; dot-dashed (red) curve, contact-interaction result; dotted (green) curve, inferred meson-cloud contribution; and dashed (blue) curve, anticipated complete result. \emph{Right} -- Pauli transition form factor, $F_{2}^{\ast}(x)$, with same legend. Data in both panels: circles (blue)~\cite{Aznauryan:2009mx}; triangle (gold)~\cite{Dugger:2009pn}; squares (purple)~\cite{Mokeev:2012vsa}; and star (green)~\cite{Agashe:2014kda}.}
\vspace*{-0.50cm}
\end{figure}

The transition form factors are displayed in Fig.~\ref{fig:NucRop_v2}. The results obtained using QCD-derived propagators and vertices agree with the data on $x\gtrsim 2$. The contact-interaction result simply disagrees both quantitatively and qualitatively with the data. Therefore, experiment is evidently a sensitive tool with which to chart the nature of the quark-quark interaction and hence discriminate between competing theoretical hypotheses.

The mismatch between the DSE predictions and data on $x\lesssim 2$ is due to meson-cloud contributions that are expected to be important on this domain. An inferred form of that contribution is provided by the dotted (green) curves in Fig.~\ref{fig:NucRop_v2}. These curves have fallen to just 20\% of their maximum value by $x=2$ and vanish rapidly thereafter so that the DSE predictions alone remain as the explanation of the data. Importantly, the existence of a zero in $F_{2}^{\ast}$ is not influenced by meson-cloud effects, 
although its precise location is.

If one supposes that $s$-quark contributions to the nucleon-Roper transitions are negligible, as is the case for nucleon elastic form factors, and assumes isospin symmetry, then a flavour separation of the transition form factors is accomplished by combining results for the $\gamma\,p\to\, R^+$ and $\gamma\,n\to R^0$ transitions: $F_{1(2),u}^{\ast} = 2 F_{1(2)}^{\ast,p} + F_{1(2)}^{\ast,n}$ and $F_{1(2),d}^{\ast} = 2 F_{1(2)}^{\ast,n} + F_{1(2)}^{\ast,p}$; where $p$ and $n$ are superscripts that indicate, respectively, the charged and neutral nucleon-Roper reactions. Our conventions are that $F_{1(2),u}^{\ast}$ and $F_{1(2),d}^{\ast}$ refer to the $u$- and $d$-quark contributions to the equivalent Dirac (Pauli) form factors of the $\gamma p\to R^+$ reaction, and the results are normalized such that the \emph{elastic} Dirac form factors of the proton and charged-Roper yield $F_{1u}(Q^2=0)=2$, $F_{1d}(Q^2=0)=1$, thereby ensuring that these functions count $u$- and $d$-quark content in the bound-states.

\begin{figure}[t]
\centerline{%
\includegraphics[clip,width=0.45\textwidth]{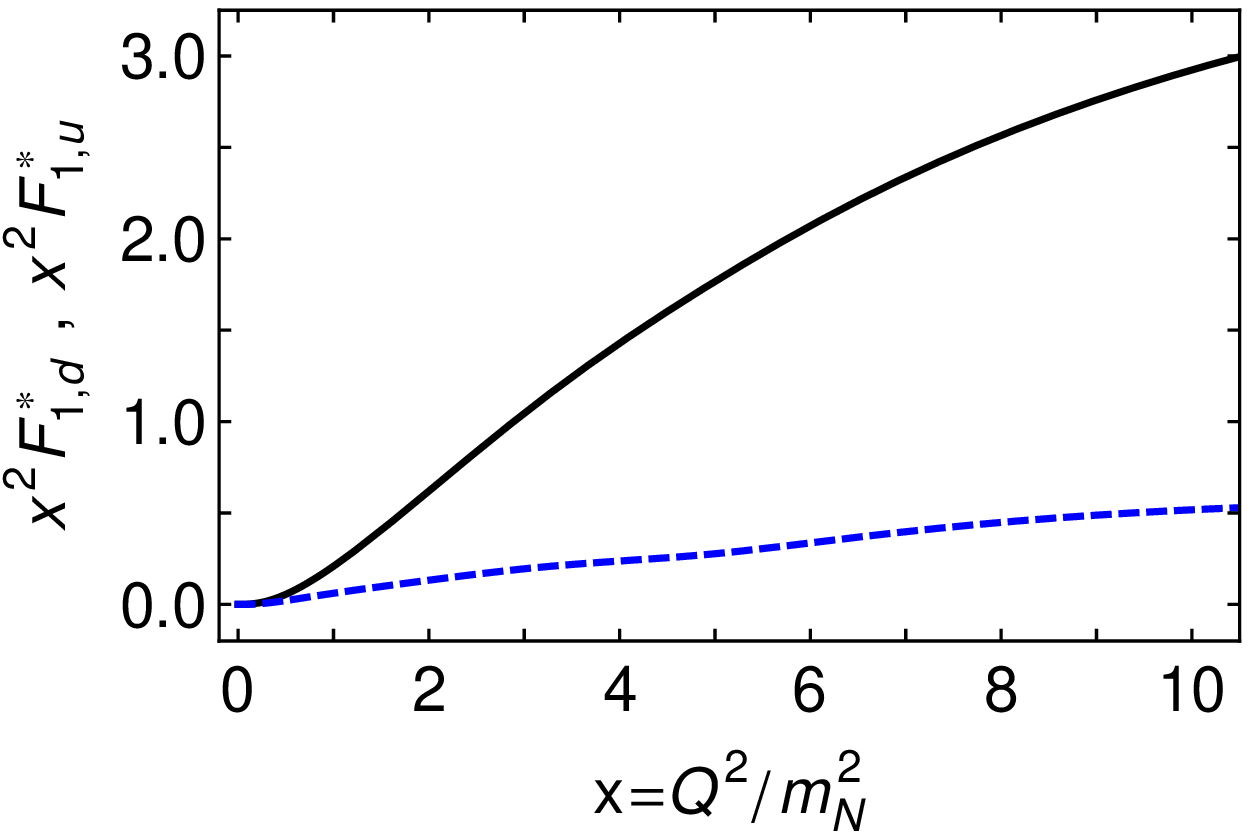}
\hspace*{0.50cm}
\includegraphics[clip,width=0.45\textwidth]{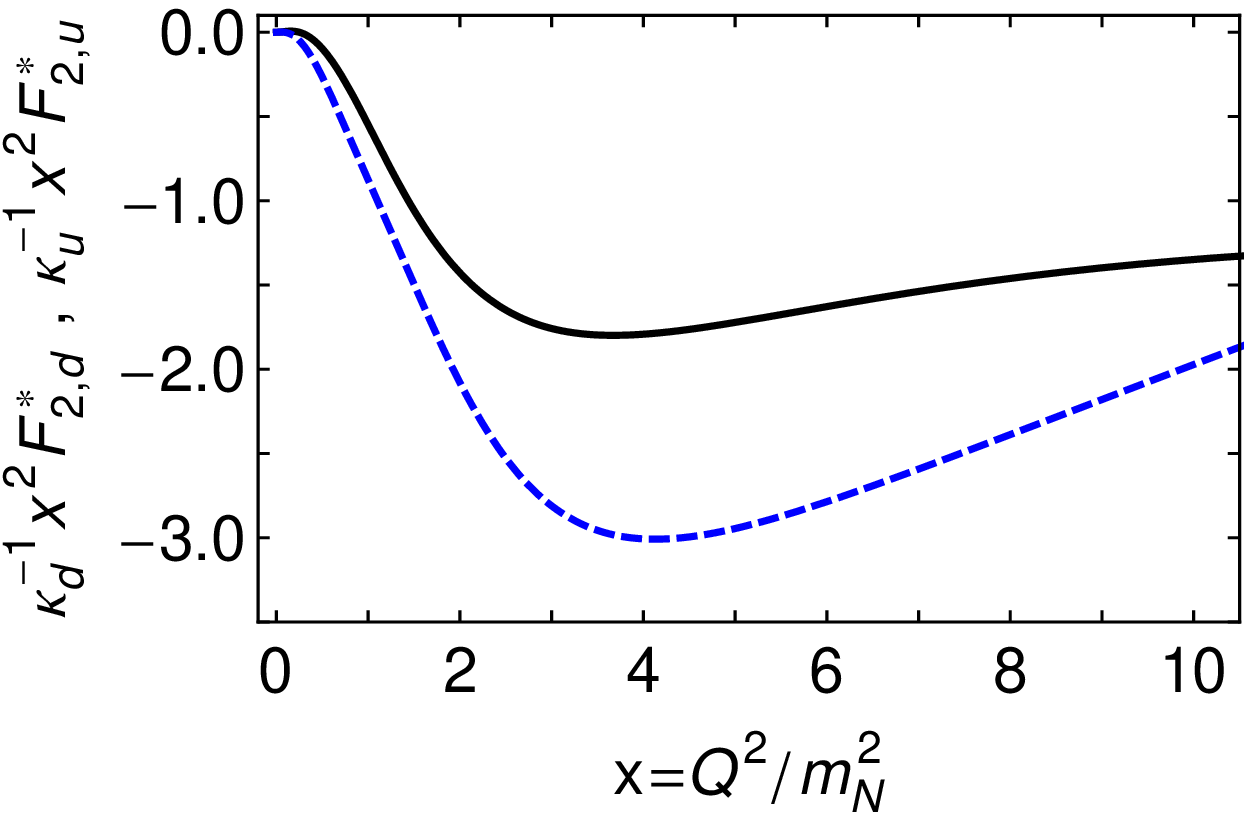}
}
\caption{\label{figxFud} $x^2=(Q^2/m_N^2)^2$-weighted behaviour of the flavour separated $\gamma\,p\to R^+$ transition form factors: $u$-quark, solid black; and $d$-quark, dashed blue. \emph{Left (right) panel} -- Dirac (Pauli) transition form factor.
\vspace*{-0.50cm}
}
\end{figure}

The left panel of Fig.~\ref{figxFud}, depicting the flavour-separated Dirac transition form factor, show an obvious similarity to the analogous form factor determined in elastic scattering~\cite{Segovia:2015ufa}: the $d$-quark contribution is less-than half the $u$-quark contribution for momenta sufficiently far outside the neighborhood of $x_N=0$ within which they both vanish; and the $d$-quark contribution falls more rapidly after their almost coincident maxima.  The noticeable difference, however, is the absence of a zero in $F_{1,d}^{\ast}$, which is a salient feature of the analogous proton elastic form factor.

The right panel of Fig.~\ref{figxFud} depict the flavour-separated Pauli transition form factor. In this instance the similarities are less obvious, but they are revealed once one recognizes that the rescaling factors satisfy $|\kappa_d^\ast/\kappa_u^\ast | < \tfrac{1}{6}$ \emph{cf}.\ a value of $\sim \tfrac{2}{5}$ in the elastic case~\cite{Segovia:2014aza}. Accounting for this, the behaviour of the $u$- and $d$-quark contributions to the charged-Roper Pauli transition form factor are comparable with the kindred contributions to the elastic form factor, especially insofar as the $d$-quark contribution falls dramatically on $x\gtrsim 4$ whereas the $u$-quark contribution evolves more slowly.

An explanation for the pattern of behaviour in Fig~\ref{figxFud} is much the same as that for the analogous proton elastic form factors~\cite{Segovia:2015ufa} because the diquark content of the proton and its first radial excitation are almost identical. In both systems, the dominant piece of the associated Faddeev wave functions is $\psi_0$, namely a $u$-quark in tandem with a $[ud]$ (scalar diquark) correlation, which produces 62\% of each bound-state's normalization~\cite{Segovia:2015hra}. If $\psi_0$ were the sole component in both the proton and charged-Roper, then $\gamma$--$d$-quark interactions would receive a $1/x_N$ suppression on $x_N>1$, because the $d$-quark is sequestered in a soft correlation, whereas a spectator $u$-quark is always available to participate in a hard interaction. At large $x_N$, therefore, scalar diquark dominance leads one to expect $F^\ast_d \sim F^\ast_u/x_N$.  Naturally, precise details of this $x_N$-dependence are influenced by the presence of pseudovector diquark correlations in the initial and final states.


\vspace*{-0.50cm}
\section{Conclusions}
\label{sec:conclusions}

We have presented a unified study of $\gamma^{\ast}N\to \Delta(1232)$ and $\gamma^{\ast}N\to N(1440)$ form factors, and compare predictions made using a framework built upon a Faddeev equation kernel and interaction vertices that possess QCD-like momentum dependence with results obtained using a symmetry-preserving treatment of a vector$\,\otimes\,$vector contact-interaction. The comparison emphasises that experiment is sensitive to the momentum dependence of the running coupling and masses in QCD and highlights that the key to describing hadron properties is a veracious expression of dynamical chiral symmetry breaking in the bound-state problem.


\vspace*{-0.20cm}
\begin{acknowledgements}
Work supported by: European Union's Horizon 2020 research and innovation programme under the Marie Sk\l{}odowska-Curie grant agreement no. 665919; Spanish MINECO's Juan de la Cierva-Incorporaci\'on programme with grant agreement no. IJCI-2016-30028; and Spanish Ministerio de Econom\'ia, Industria y Competitividad under contract nos. FPA2014-55613-P and SEV-2016-0588.
\end{acknowledgements}


\vspace*{-0.50cm}

%
%
%

\end{document}